\documentclass[runningheads]{llncs}
\usepackage{graphicx}
\graphicspath{ {img/} }
\usepackage{float}
\usepackage{arabtex}
\usepackage{utf8}
\usepackage{appendix}
\usepackage{amsmath}
\newcounter{magicrownumbers}
\newcommand\rownumber{\stepcounter{magicrownumbers}\arabic{magicrownumbers}}

\begin{document}
\title{Arabs and Atheism: Religious Discussions in the Arab Twittersphere}
\author{Youssef Al Hariri \and Walid Magdy \and Maria Wolters}
\authorrunning{Y. Al Hariri, W. Magdy and M. Wolters}
\institute{School of Informatics, The University of Edinburgh, Edinburgh, UK \\
\email{\{y.alhariri,wmagdy,maria.wolters\}@ed.ac.uk}\\
}
\setcode{utf8}
\maketitle     

\begin{abstract}
Most previous research on online discussions of atheism has focused on atheism within a  Christian context. In contrast, discussions about atheism in the Arab world and from Islamic background are relatively poorly studied. An added complication is that open atheism is against the law in some Arab countries, which may further restrict atheist activity on social media. In this work, we explore atheistic discussion in the Arab Twittersphere. We identify four relevant categories of Twitter users according to the content they post: atheistic, theistic, tanweeri (religious renewal), and other. We characterise the typical content posted by these four sets of users and their social networks, paying particular attention to the topics discussed and the interaction among them.
Our findings have implication for the study of religious and spiritual discourse on social media and provide a better cross-cultural understanding of relevant aspects.

\keywords{Atheism \and Arabic \and Religion \and Islam \and Twitter }
\end{abstract}
\section{Introduction}\label{introduction}

Several reports show that the number of Arab atheists is growing\footnote{Sources: https://newhumanist.org.uk/articles/4898/the-rise-of-arab-atheism\\
https://www.washingtontimes.com/news/2017/aug/1/atheists-in-muslim-world-growing-silent-minority/} in a region that is often intolerant to atheists~\cite{BenchemsiAhmed2015Iaos,PatrickKingsleyinCairo2014Ehth,nabeel_2017,HunterStuart2016THLo}.
The limited existing studies analysing discussions about atheism online~\cite{chen2014us,ritter2014happy} focused on Western and Christian societies. 

In this work, we investigate how atheists in the Arab societies leverage Twitter to discuss their disengagement from religion, mainly Islam, which is the dominant religion in the Arabic region, and the interactions by different users around this topic.
We characterise relevant types of user accounts by distinguishing between four main groups: Arab atheists, who do not believe in a deity; Arab theists, who believe in a religion; Arab Tanweeri's, who believe in Islam but also accept other beliefs and promote religious reform; and Other, who do not openly discuss their religious views.
We focus on the main topics discussed online by each of these user groups, and the way in which their opponents engage with them through replies and retweets.

For our analysis, we collected and analysed the tweet timelines of around 450 user accounts who were heavily engaged in discussions concerning atheism. Our study investigates the following research questions:
\begin{itemize}
\item{RQ1:} What are the common topics and features that Arab Atheists share that distinguish them from  Theists, Tanweeris, and Others?
\item{RQ2:} How do Arab Atheists interact with the other three groups?
\end{itemize}

Our analysis shows that there are active online discussions between Arabs from across the religious spectrum.
Most of the discussions are related to local and regional topics and mainly topics related to the oppression of women.
The vast majority of Arabs who believe in a deity are from Saudi Arabia and they show solidarity with their government.
Arab Christians are more likely to argue against Islam than to argue against Atheism.

We believe that this study sheds the light on one of the most interesting and sensitive topics in the Arab world that has been relatively neglected in the literature. Our findings should promote research in this direction to have further in-depth analysis to the topic of atheism and religion in the Arab world.

\section{Background}
\label{sec_background}
\subsection{Religions vs. Atheism, Globally and in the Arab Societies}

Religion is still a force to be reckoned with in today's world \cite{1999Tdot}. 
Both religiosity and non-religiosity are highly complex and multifaceted. In this paper, we will adopt concepts that are particularly well suited to the context of discussing Islam. 
In general, religions consist of a community of believers who share tenets of faith and practices of worship, some of which may require separation from others \cite{croucher_zeng_rahmani_sommier_2017,durkheim1976elementary,reesReligionCulture2018}. Within a religion, there are often many branches which may or may not coexist peacefully with each other. For example, Islam has two major denominations, Sunni and Shia, which in turn are split into many branches and sub-denominations \cite{Shahrastam1984Msad}. 

Non-religiosity is also highly diverse. It  includes wider groups such as humanism, indifferentism, secularism, agnosticism, irreligion, anti-religions and atheism \cite{2017Bge,QuackSchuh2017}. Around  16\% of the world's population identify as atheist or non-religious \cite{GRD2018}. 
While atheists are minority in Arab countries \cite{banks2017citizenship,GRD2018,GRL2013Tgrl},
the number of Arab atheists appears to have increased noticeably recently despite the harsh penalties for atheism in several Arab countries \cite{BenchemsiAhmed2015Iaos,RCC20102018,HunterStuart2016THLo}. 
A poll conducted in 2012 shows that an average of 22\% of Arabs expresses atheist views, or at least some measure of religious doubts by using their social media accounts \cite{BenchemsiAhmed2015Iaos,RCC20102018}.

\subsection{Social Media, Globally and in the Arab Societies}

Twitter has 336 million active users per month \cite{statistic_272014} and its data has been used before to study religions on social media. In \cite{chen2014us} the authors analysed more than 250k Twitter accounts to understand the main features of religiosity on Twitter for users from the US.
The work found a reasonable positive correlation between Twitter data, i.e. declared religions, and offline surveys data for geographic distribution of religious people.
The study includes analysing the tweets and networks for each user to identify discriminative features of each religious group and to study the linkage preference.
It shows that the networks dynamics, mainly followers, friends, retweets and mentions, tell more about the religious users and provide more effective features than the tweets contents. 
They also observe that Twitter users tend to interact more with users within the same religion.

Arabs have positive views about social media and its influence on their societies.
According to \cite{RachaFadi2012}, Arabs are influenced positively towards other cultures, opinions, views and religions after their involvement in social media. Social media has many different functions. It facilitates the revolutions spread during the Arab Spring in 2011 \cite{RachaFadi2012}, but it also serves as a propaganda and recruitment venue for  extremist groups around the world \cite{AwanImran2017CIat,PoT426,MagdyDarwishandWeber2016,RicheyBinz2015}. It also serves as a platform for underrepresented groups, such as Arab atheists, to communicate and show their existence. 

There is surprisingly little work on online atheist communities within the Arab or Muslim societies. A notable exception is  \cite{SaskiaSchäfer2016FFIO}. In her study of Indonesian atheists, she found that social media helped atheists activists to safely highlight their existence in a religious country, Indonesia, show their positive side, and build a thriving community. 
However, they risked exposure through contact with human rights activists around the world through social media.

Far more attention has been paid to radical theists in the Arab world. Magdy et al. \cite{MagdyDarwishandWeber2016} sought to understand  the origins and motivations of ISIS Arab supporters by comparing data for about 57,000 Arab Twitter accounts before and after the emergence of ISIS. They find that historical data can be used to train a classifier to predict a user's future position on ISIS with an  average F1-score of 87\%.
There are also clear differences in the topics discussed.
ISIS opponents are linked to the position of Arab regimes, rebel groups, and Shia sects; while ISIS supporters talk more about the failed Arab Spring. \cite{MagdyDarwishandWeber2016}.
Interestinly, the most widely used distinctive hashtag used by ISIS supporters  ``\#Million\_Atheist\_Arab'' which was part of a campaign by Arab atheists. This indicates that the topic of atheism is well known and discussed in the Arab world, despite the lack of studies.

A particular facet of religious discussion on  Arab social media is hate speech about both religion and atheism \cite{Albadi2018hate}. The study by Albadi et al. shows that 42\% of the studied tweets ($n$=6000) that cross-reference religions contain hate-speech.

We conclude that there is a clear gap in our knowledge of religious discourse and dialogue between atheists and theists on the Arab social media. We propose to address this gap by applying a quantitative analysis to characterise the Arab online theist and atheist communities.

\section{Data Collection} \label{sec:data_collection}

\subsection{Collecting Active Users Discussing Atheism}
For retrieving relevant accounts, we received a list of 200 Arabic Twitter accounts from Bridge Foundation\footnote{\url{https://bridges-foundation.org/}}, a non-profit organisation based in London that aims to build bridges between Islam and other religions. These accounts had been labelled by their volunteers as promoting atheism content.
We manually reviewed these accounts by inspecting their description and shared content. We only kept those that 1) explicitly mentioned that they are atheists and 2) promote atheism or clearly criticise religions in the majority of their tweets. Thus, we ended up with only 80 accounts that met our criteria. We used these 80 accounts as our seeding accounts.
We used the Twitter streaming API to collect all the tweets that interacted with these accounts for 4 months between Feb and May 2018.  We collected over 100K tweets during that period and limited our analysis to those 434 user accounts  interacted with the seed accounts over 200 times, either by retweeting, replying, or mentioning them. We consider these to be the most active users on the topic of atheism on Arab social media at that time. 
For these 434 accounts, we collected their entire Twitter timeline to study their network interactions and the content they discuss in their tweets. In total, we collected a set of 1.3M tweets for these accounts.

\subsection{Data Annotation}
After careful inspection of the data, we labelled the accounts based on the contents of their tweets and the beliefs they promote in their timelines.  The four labels we devised according to the content are: 
\begin{itemize}
    \item \textbf{Atheistic} content that promotes content denying the existence of God (or gods) or explicitly rejects a religion (or religions) without any sign of religious affiliation.
    \item \textbf{Theistic} content that shows belief in God or disclose a religious affiliation and defends it.

    \item \textbf{Tanweeri} content that shows affiliation to Islam (or another religion), but promotes religious reform and accepts other beliefs.
    \item \textbf{Other} none of the above.
\end{itemize} 

Three native Arabic speakers, from three different Arab countries, received a workshop training for labelling the accounts. The main purpose of the training was to ensure clear understanding of the annotation guidelines and isolating any personal beliefs while annotating the data. 
Each annotator was instructed to inspect most of the collected tweets for each of the accounts before making a judgement. They also had access to the user description and link to their online profile to assist them making decisions if needed.
Judgements were based both on the Twitter users' own tweets and on frequent retweets of a particular stance.

Initially, 50 accounts were labelled by all three annotators. Cohen's Kappa values between each annotator pair were 0.732, 0.592, and 0.634, which reflects the subjective nature of interpreting statements of belief. The main confusion was found between the `Atheistic' and `Tanweeri' labels. After discussion of sources of disagreement, the annotators proceeded to label the remaining 383 accounts. The average time for labelling one account ranged between 15-30 minutes.

In addition, the annotators tried to identify if the account belongs to an individual person or a formal entity that promotes certain stances. For individual accounts, we also recorded gender, if it was identifiable. 

\begin{figure}[t]
  \centering
      \includegraphics[width=\linewidth]{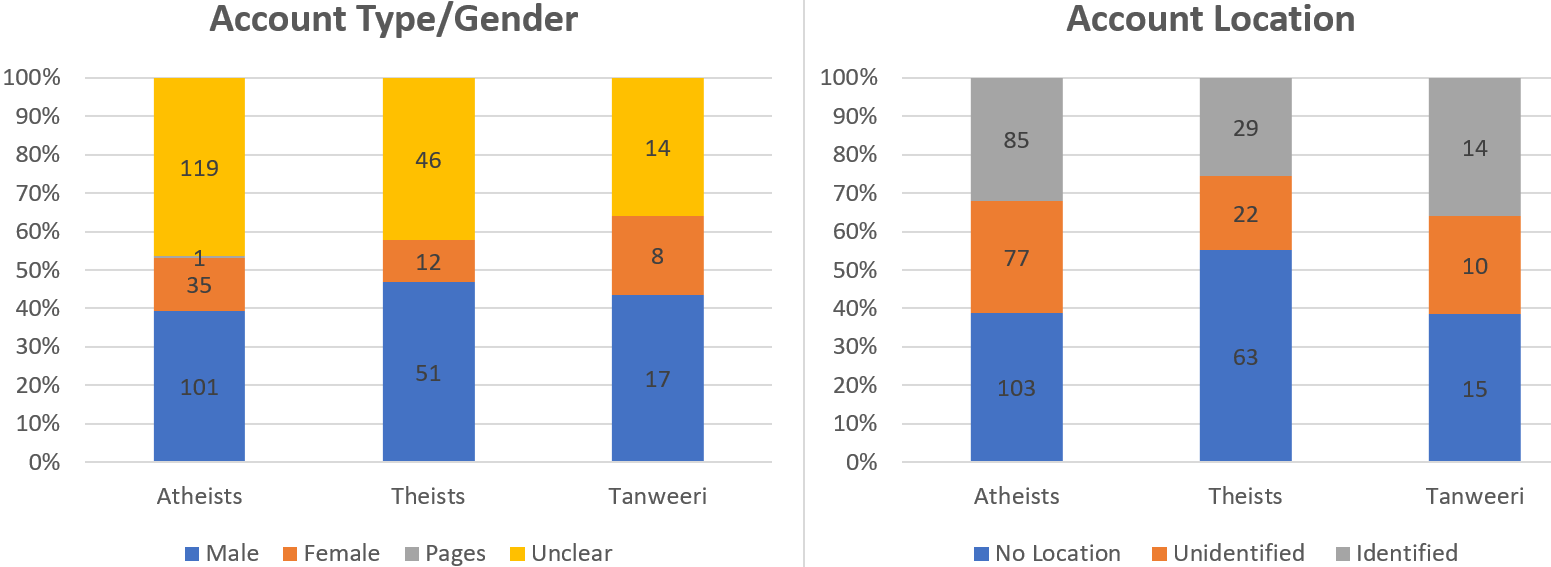}
  \label{fig:distributionOverClasses}
  \vspace{-0.5cm}
  \caption{Type, Gender, and Location of Accounts}
\end{figure}

\subsection{Data Statistics}

Since the seed list consisted of atheist accounts, it is not surprising that most of the accounts we labelled belong to atheistic class ($N=256, 59\%$). 109 accounts ($25\%$) are labelled as theistic, 39 ($9\%$) as tanweeri, and 30 ($7\%$) as Other. 

Regarding the type of the accounts, only one atheist account is a page, while all the others are personal accounts for Twitter users. Figure \ref{fig:distributionOverClasses} shows the distribution of accounts gender for each of the categories. In addition, it shows which of these accounts has an identifiable location listed in their profile. As shown, there are more males than females in the collected accounts, while for around 40\% of the accounts gender was not clear. Tanweeri users have the largest percentage of females among other groups and they tend to declare their gender more than other groups. 
For location, theistic users have more identifiable locations than the atheists and tanweeris. This might be for security reasons of atheists to protect themselves against laws in some of the Arab countries. Most of the identifiable locations are from Saudi Arabia for all accounts, followed by USA for atheist and theist accounts. Details on identified location of users in each group is shown in Appendix in Figure \ref{fig:locationsData}.

\section{Analysis of Atheism Discussions on Twitter}

For each account, we apply our analysis on both the content and the network interactions of the accounts. This includes the accounts they retweet their tweets (\textbf{retweets}), the accounts they reply to (\textbf{replies}), the accounts they mention in their tweets  (\textbf{mentions)}, the hashtags used by the account (\textbf{hashtags}), and the Web domains linked in their tweets  (\textbf{domains}). In the following discussion, all tweets are rephrased to protect the original posters. 

\subsection{Top Discussed Topics}

In this section, we analyse the topics that are most frequently discussed within the 1.3M tweets in the timelines of all the 434 accounts. This should highlight the discussion topics by the most active Arab users on atheism.
We used hashtags to describe the topics that users talked about. Tweets will not be quoted, so that they cannot be traced back to their authors. However, we provide example English translations of those tweets.

As shown in Table \ref{tab:tophashtagssmall} (more details in Appendix Table \ref{tbl:allfreqhastags}), almost all classes talk about similar topics such as rationalists, Middle-East countries, ISIS, women's oppression, Saudi women's rights, regional conflicts, and topics related to atheism and reformation, such as tanweer, ex-Muslims, atheists and atheism. Table~\ref{tab:hashtagsTable} gives few example tweets of the usage of these hashtags in context.

The hashtag CreatingAlmohawer (training the interlocutor) refers to an online program designed to train Muslims to rebut unfounded claims about Islam.
Most of Arab atheists do not only argue against Islam, but against all religion, specifically the Abrahamic religions (Judaism, Christianity and Islam). However, Islam is the most discussed religion as most of them were Muslims. Relevant  hashtags include Abrahamic dice and Former\_scriptures; relevant tweets argue that since archaeology provides counterevidence to the Jewish Bible, this makes all Abrahamic religions invalid. 
Other tweets questioned the existence of Moses, a prophet in Abrahamic religions, and claimed that there is no evidence to prove his existence.
Interestingly, the hashtag Former\_scriptures is also used by some non-Muslim theists  to argue against Islam.

\begin{table}[h]
    \centering
    \begin{tabular}{|p{1\textwidth}|}
    \hline
         \textbf{All}: \scriptsize Rationalists, SaudiArabia, CreatingAlmohawer (interlocutor training), ISIS, Islam, StephenHawking, Israel, Egypt, Iran, Truth, Yemen, Syria, Friday, CEDAWSaudi, SaudiWomenDemandDroppingGuardianship, Qatar, Kuwait, DroppingGuardianship, AForgottenWomenPrisoners \\ \hline
         \textbf{Atheistic}: \scriptsize DelusionTrade, ExMuslim, Atheism\textsuperscript{*}, EvolutionFact, Atheist\textsuperscript{*}, Tunisie, SaveDinaAli, TweetAPicture, RaifBadawy, Science, BlessedFriday, QuranInPictures, Trump, YouthTalk, Sweden, WomenCarDriving, FreeSherifGaber, Woman, ALogicalQuestion, DontSayIAmDisbeliever, WhereIsAminah, OsamaAljamaa, UnveilingIsNotMoralBreakdown, WomenInternationalDay, ViolenceAgainstSaudiWomen\\ \hline
         \textbf{Theistic}: \scriptsize Atheist, Atheism, ChildrensMassacreInAfghanistan, Pray, SpreadOfIslam, ChristianityFact, Jesus, Palestine, Bible, AlAzharIsComing, Quran, Atheists, LegalizationOfZionization, Urgent, Continued, DefenceQuranAndSunnahByProofs, AssadBombardDomaChemicalWeapons, Christianity, MesharyAlAradah, Jesus, NaizakTranslation, Gaza, Aleppo, Turkey, IranProtests \\ \hline
         \textbf{Tanweeri} \scriptsize HashemiteOccupation, OAyedDoNotSteal, WeakHadithEmployedBySahwa, CleaningSchoolsFromSururiWomen, FlutesRevelation, CrownPrince, Al-NassrFc, CrownPrinceOnCBC, SlaveryAllowanceForSaudiWomen, HowISurviveFromSahwa, RefusedToReleaseHisDaughter, IDecidedToWearItOnMyHead, AlmutlaqAbayaIsNotObligatory, Brothers, Yemen, SaudiWomenProudOfGuardianship, SaveMeFromViolence, CompassWithIslamBahiri, NoClouserShopsDuringPrayTime, MyFaceVeilIsHonor, SaudiCinema, MajidaElRoumi, CinemaInSaudiArabia, OffendedWomenOnly-GymClosed, MBSInterviewsTheAtlantic \\ \hline
    \end{tabular}
    \caption{Top 25 hashtags from each class translated into English. Full details of hashtags shown in Appendix}
    \label{tab:tophashtagssmall}
\end{table}

\begin{table}
    \centering
    \begin{tabular}{|c|c|p{\textwidth}|} 
    \hline
\scriptsize \rownumber & \scriptsize A & \scriptsize Religious drugs generate huge profits for delusion traders and dealers and more poverty for the people \#Rationalist. \\ \hline
\scriptsize \rownumber & \scriptsize A & \scriptsize \#Delusion\_traders successfully make simple minds fools, mindless and inhuman. \\ \hline
\scriptsize \rownumber & \scriptsize A & \scriptsize I am going to publish a simple introduction to Palaeontology Which is an overwhelming proof to the \#EvolutionFact; Follow me. \\ \hline
\scriptsize \rownumber & \scriptsize A & \scriptsize Anyone benefited from the diffusion would promote myths and delusions \#DelusionTrade. \\ \hline
\scriptsize \rownumber & \scriptsize A & \scriptsize @user: \#ViolenceAgainstSaudiWomen \#SaveDinaAli where is Dina Ali? she was disappeared since a year. \\ \hline
\scriptsize \rownumber & \scriptsize A & \scriptsize @hrw\_ar you need to prove credibility to protect that girl from being killed by her family. \\ \hline
\scriptsize \rownumber & \scriptsize A & \scriptsize @user: \#ViolenceAgainstSaudiWomen Religions shouldn't be a law ..  \\ \hline
\scriptsize \rownumber & \scriptsize A & \scriptsize We are living in 2018 and still there are people being arrested for expressing their political views and religious beliefs. We should have legal codes to protect the freedom of speech. \#freeSherifGaber \#FreeSherifGaber  \\ \hline
\scriptsize \rownumber & \scriptsize A & \scriptsize \#ViolenceAgainstSaudiWomen \#SaudiWomenDemandDropGuardian640 \#StopEnslavingSaudiWomen We demand justice for female victims of domestic violence. \\ \hline
\scriptsize \rownumber & \scriptsize T &\scriptsize A A Darwinian Atheist says please help Dina. Why we don’t consider her story as a natural selection or an evolutionary development? \#SaveDinaAli  \\ \hline
\scriptsize \rownumber & \scriptsize T & \scriptsize Atheists did not support Muslims to liberate lands or to defend themselves, but they believe that they have the right to live between them. \#rationalists  \\ \hline
\scriptsize \rownumber & \scriptsize T & \scriptsize The Gravity theory Scientist believes in God and says atheists are the most stupid. \\ \hline
\scriptsize \rownumber & \scriptsize T &\scriptsize Some atheists talk about the capital punishments for atheists in Islam; However, they ignore that it is applied through a justice body. \#rationalists  \\ \hline
\scriptsize \rownumber & \scriptsize T &\scriptsize \#FreeSherifGaber This is the penalty for any beggar who trades in atheism and asks for funding to produce rotten mould. \\ \hline
\scriptsize \rownumber & \scriptsize T &\scriptsize \#FreeSherifGaber he worked for months to prepare a storytelling with full of lies, ignorance and fabrication but the response is quickly found. \\ \hline
\scriptsize \rownumber & \scriptsize T &\scriptsize RT @user Anyone claims that violence against women and children is allowed in Islam is a liar. \#Al-AzharIsComing\\ \hline
\scriptsize \rownumber & \scriptsize T &\scriptsize @hrw\_ar We will stay protected by our families and you should stop attacking our religious and cultural heritages. It is a crime against us. \\ \hline
\scriptsize \rownumber & \scriptsize T &\scriptsize \#ChildrensMassacreInAfghanistan USA has problem with the Holy Quran not with Muslims. \#Rationalists\\ \hline
\scriptsize \rownumber & \scriptsize W &\scriptsize Yes, the cost is prohibitive; there will be mass destruction, killing and displacement. But it is less costly than governing the Iranian criminal gangs \#HashemiteOccupation\\ \hline
\scriptsize \rownumber & \scriptsize W &\scriptsize \#OAyedDoNotSteal, what are we did not discover yet from the Sahwa era?.\\ \hline
\scriptsize \rownumber & \scriptsize W &\scriptsize Lots of Hadiths were fabricated by Sahwa scholars and it is time to execute them.
WeakHadithEmployedBySahwa\\ \hline
\scriptsize \rownumber & \scriptsize W &\scriptsize \#WeakHadithEmployedBySahwa Assassinating \#Sahwa is a national duty. \\ \hline
\scriptsize \rownumber & \scriptsize W &\scriptsize \#CleaningSchoolsFromSururiWomen
The school is an educational body. It shouldn't be part of a religious party and it is unacceptable to be used for the interests of some!\\ \hline
\scriptsize \rownumber & \scriptsize W &\scriptsize Schools are the most places to spread Sahwa thoughts specifically women teachers of schools in Riyadh. \#CleaningSchoolsFromSururiWomen\\ \hline
\scriptsize \rownumber & \scriptsize W &\scriptsize \#CleaningSchoolsFromSururiWomen Obligating students to wear veil with the face cover enforces them to follow a certain jurisprudential.\\ \hline
\scriptsize \rownumber & \scriptsize W &\scriptsize Soon women will travel and enjoy their full rights the same as men. May Allah prolong the life of this leader. \#CrownPrinceOnCBC\\ \hline
\scriptsize \rownumber & \scriptsize W &\scriptsize Soon there will be Shia members in the Council of Ministers and in the government. Also, the president of the most important university in KSA is Shiite.
We have a mix of Islamic schools and sects \#MBSInterviewsTheAtlantic\\ \hline
    \end{tabular}
    \caption{Sample (translated) tweets with Significant Hashtags.\scriptsize A: Atheistic, T: Theistic, and W: Tanweeri timeline}
    \label{tab:hashtagsTable}
\end{table}

Topics related to the oppression of women attract users from all groups. The hashtags `SaudiWomenDemandDroppingGuardianship' and  `StopEnslavingSaudiWomen'  come from a long-term online campaign led by Saudi women who want freedom from social restrictions and supported by feminists from the region and around the world.
Atheist users claim that the Saudi women must have their freedom of choice without the guardian system. For example, they will explicitly demand dropping restrictions on travelling, obtaining a passport, and driving. While some theist users show sympathy with the cause, most of them reject it. For example, one tweet claims that travelling on a passport is already possible with the guardian's electronic permission, and another claims that the campaign for dropping guardianship is managed by men. Similar polarisation are found around the hashtags talk about different cases of oppression against women such as WhereIsAmna, AbusedWomanInAbha, SaveDinaAli, RefusedToReleaseHisDaughter and MajedManaOmairOppresseHisWife. All of these hashtags are related to cases of women in Saudi Arabia. Tweets number 5, 7, 9 and 10 in Table \ref{tab:hashtagsTable} shows some tweets related to these hashtags.  Some theists support the victim, but others try to find excuse for the case. For instance, some theists in the latter hashtag claim that the wife is benefiting from the accusations and accuse her of treason. 

All groups intensively discussed terrorism and terrorist groups.
While atheists blame Islam for terrorist groups, such as ISIS, 
theists claim that ISIS is used by Islamophobes to equate Islam with terror.
It is noticeable to see that Atheists prefer to mention ISIS by using its English acronym, Arabic name (The Islamic State \<الدولة الإسلامية>), or the acronym Da'esh ( \<داعش>) within hashtags such as Da'esh\_is\_an\_Islamic\_Product.
On the other hand, theists prefer to mention ISIS by its Arabic acronym.
While some Tanweeris argue that ISIS is not the real Islam, others blame the religion for the spread of terrorist groups.

In addition, the data set shows interactions with international organisations conducted from both atheists and theists.
Atheists are willing to contact international organisations to seek protection or to promote their opinions (example: Tweet 6, Table~\ref{tab:hashtagsTable}), while theists also actively discuss their point of view on similar topics. 

For instance, in Spring 2017, Human Rights Watch (HRW) tweeted that
``An emergency case resulted from guardianship law in KSA \#SaveDinaAli''. 
A theistic account denied that and argued that she might have escaped after committing a crime or there is a hidden information.
Another tweet published by HRW argues that while allowing women to drive is a step forward, but the guardianship law in Saudi should be abolished. A female theist replied with tweet number 17 in Table \ref{tab:hashtagsTable}. Another theist wrote:
``@hrw\_ar it is not your business''.
In another tweet, HRW quoted a claim of prisoner abuse in Saudi Arabia that was made by the  New York Times. A theist denied that news and wrote: ``You should have truth, most of these news are fabricated''.
These samples show that Arab theist society actively engages with reports by other organisations.

Theists are more likely to talk about Arab countries and Middle East countries, such as Iran and Turkey, than atheists and tanweeris, whereas these two groups are more interested in topics relating to Saudi Arabia.
All classes are divided in their opinion about conflict regions in the Middle East. 
For example, the majority of atheists look forward to dramatic changes in relationships of Arab countries with Israel while some of them refuse any rapprochement.
However, most of theists claim that news of such changes, especially regarding the relationship between Israel and Saudi Arabia, are fabricated.

\subsection{Distinctive Topic Discussion by Groups}

To have a clear understanding about the trends that are mostly used, we ran a logistic  regression analysis for each of the three main groups (Atheist, Theist, Tanweeri) with the top 50 hash tags as features. The hashtags with the highest weight are considered to be particularly distinctive.
A sample of the top hashtags used by Atheists, Theists and Tanweer groups are shown in the Appendix table \ref{tbl:aafreqhastags}. Notice that the table shows the frequencies of each hashtag from the three groups.

The most frequent hashtags mentioned by atheists are related to evolution theory, delusion trade, atheism and leaving Islam. 
The data set shows that Arab atheists strongly support evolution theory, and provide evidence to convince others. 
Most theists are not interested in discussing the theory, while others respond with the hashtag (\<التطور>\_\<خرافة> - ``Evolution is a myth''). 
Atheists also show solidarity with other atheists or activists.
That is clear from hashtags such as FreeSherifGaber, RaifBadawy, AbdullahAlQasimi (one of the most controversial Saudi writers), and OsamaAljamaa (Saudi Psychologist), which are cited and retweeted by atheists. Abdullah Al Qasimi, as described in tweets, changed his position from being an Islamic Salafi scholar to defending atheism and tanweer, while Aljamaa has written about personal development and self-awareness, and his works are cited widely by Arab atheists. Finally, atheists talk more about atheism and leaving Islam. Relevant hashtags include ExMuslim, Atheism, Atheist, TheReasonWhyILeftIslam, and ExMuslimBecause.

Table \ref{tbl:aafreqhastags} shows samples of the 15 most frequent hashtags that theists used in their tweets. 
Most of the accounts in this category are located in Saudi Arabia, which can be inferred from country-specific hashtags such as MohammadBinSalman,
TurkiAlSheikhThePrideOfPeople.
Arab theists widely discuss atheism and atheist by using their Arabic names (\<إلحاد> - atheism) and (\<ملحد> - atheist).
Also, it is clear that most of theistic content either discusses or criticises Christianity.
This is shown by hashtags such as Truth about Christianity, Jesus, Christianity, Contradictions of the Bible, and Books about Christianity.
In addition, hashtags that talk about terrorism are specifically used by this class of authors.
Theists talked about ISIS by using its short Arabic name \<الدولة> (the state) and its leader (Al-Baghdadi). Theists also talk about Al-Qaeda, its Syrian branch (Al Nosra front and Hayat Tahrir Al-Sham), and their leader (Algolani and Al-Zawahiri), as well as conflict regions such as Al-Raqqah, Afghanistan, Syria, Palestine, Gaza, Aleppo, Iran and Yemen. One of the most significant hsahtags used by theists is \#Al-AzharIsComing. 
It is published by a famous Egyptian Islamic scholar.
His tweets are widely retweeted by theists, and as described in his tweets, Al-Azhar is one of the oldest academic bodies in the Islamic countries.

The typical  hashtags that are used by Tanweeris are a mix of different cultures and opinions.
One of the most discussed topics between tanweeris is the Hashemite Occupation, which talks about Islamic sects that took control over Yemen.
Some of tweets show refusal of the existence of Islam as a religion in Yemen, but most of them talk about the conflict in Yemen and rejection of Houthis.
Also, Tanweeris discussed a wider spectrum of Islamic parties and movements.
However, most of their discussions show solidarity with their governments against different Islamic parties and scholars.
In addition, they show a clear rejection of the opinions of scholars and sheikhs.
In fact these tweets are also evidence for their solidarity with the government in KSA.
This is reflected in tweets related to the Crown Prince of KSA Mohammad bin Salman interviews as shown in tweets 26 and 27.

Interestingly, the most frequent hashtags are related to Saudi football, in particular to a  club from the Capital city `Riyadh'.
Most of the accounts with tanweeri contents are fans of this club.
The hashtag Urawaian Proverbs is used by Al-Nassr FC fans to mock another team from the same city after it was defeated by Urawa Reds FC. 

\subsection{Network Interactions around Atheism}

Analysing the social network is an important step towards understanding the motivation of Arab atheists to declare their beliefs online. Here, three types of interaction network are analysed, user mentions in self-written tweets, mentions in replies, and accounts that they retweet. 
Due to potential repercussions for the Twitter users mentioned, especially since some Arab countries criminalise atheism, we will not list the names of the accounts, unless they are official news sources, but instead characterise their content. Account names are available on request from the authors after signing a confidentiality agreement.

\begin{figure}[h]
 \centering
 \includegraphics[width=1\linewidth]{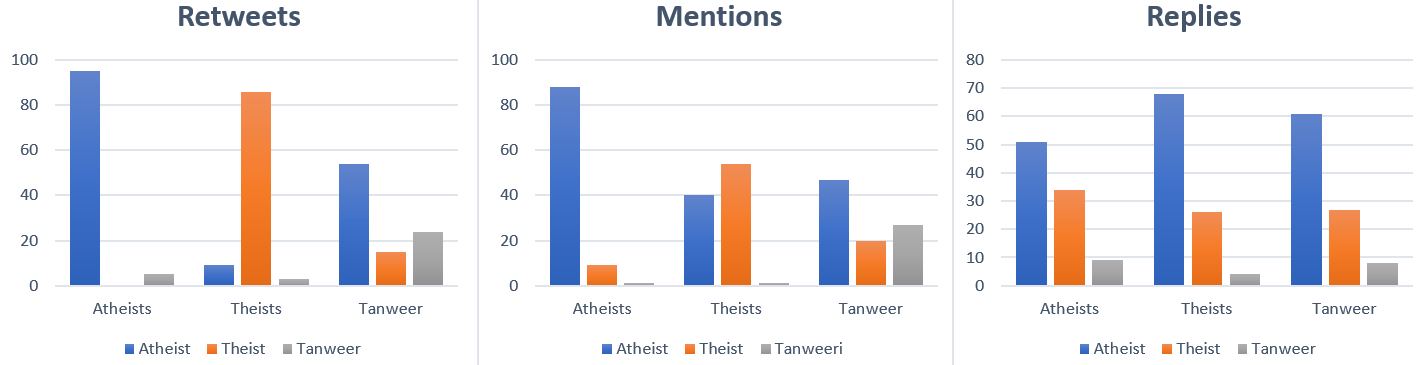}
 \vspace{-0.3cm}
 \caption{Interaction Network of Each Group}
 \label{fig:network_mention}
\end{figure}

Interestingly, as shown in Figure \ref{fig:network_mention} Atheists are more likely to mention, reply and retweet to members from their groups. Also, they are the most mentioned accounts by users of different beliefs.
This is aligned with the previous findings that Arab societies openly discuss their beliefs online.
It might be good to investigate more the tweets that both Atheists and Tanweeris reacted to, especially given that Tanweeris are less likely to be mentioned by atheists.
Atheists are also very active in publishing replies to members from all groups.
A sample of these tweets show that they support each other, defend their opinions, convince others,  and discuss others' beliefs.
Arab Theists are more likely to retweet each other than to retweet  other groups.
They amplify significant tweets, such as tweets published to explain a phenomenon and link it to religious belief or to support their opinions.

Even though many of the accounts mentioned in Atheist tweets are self-described atheists, 
the most frequent mentioned account 
belongs to a well known supporter of the measures taken by the new Saudi leadership.
The accounts mentioned most frequently by theists are mostly belong to a famous religious figure or to an active theist who defend Islam. %
Most of the Muslim believers argue against atheism and promote Islam in their timelines. There are some Christian believers who argue against Islam but not against atheism.

\subsection{Domains Analysis}\label{sec:domainanalysis}

Web domains might give information about the source of information each group prefer. 
Hence, we analyse the most frequent domains used by each class that are shown in Appendix Table \ref{tbl:freqDomains}.

The most frequent websites used in tweets by atheists are related to social media platforms such as instagram.com, facebook.com, pscp.tv, curiouscat.me and ask.fm. The reason for this might be that these websites help them link to and reach out to other atheists in their societies, and share posts with atheism relevant content. 
The domain  ``wearesaudis.net'', also frequently mentioned by atheists, is an online forum that provides suggestions and guidance on how to seek asylum in different countries including Israel. Atheists are also more interested in online resources about science, such as ibelieveinsci, and they often interact with 
non-Arabic news websites such as dw.com, bbc.com,
arabic.rt.com, theguardian.com, independent.co.uk, f24.com, dailymail.co.uk, and  nytimes.com.
Atheists also widely share online campaign posts from change.org, which hosts many human rights petitions, and the domain of the organisation Human Rights Watch,  hrw.org. This organisation covers human rights in the Middle East, especially the Arab Spring countries, and Saudi Arabia.

On the other hand, the most frequent domains used by the theists are du3a.org, d3waapp.org, alathkar.org and 7asnat.com. These sites are auto-post services for Islamic supplications, duas, and notifications.
In addition, qurani.tv and quran.ksu.edu.sa are frequently used, but we were unable to  determine if they are used as auto-post services or cited actively. 
The news sources preferred by believers are those written in Arabic, such as the Arabic service of Russia Today (arabic.rt.com), Saudi Press Agency (spa.gov.sa) and Sabq News (sabq.org). The most referenced non-arabic news source by Arab theists is cnsnews.com. However, most of theists believe that it not a trusted source; it is regared as a ''liar`` and ``a conservative and right wing American source''.

Like Atheists, Tanweeris often share content from other  online social media platforms on Twitter. Relevant URLs include curiouscat.me, instagram.com, facebook.com, and pscp.tv. They also uses the tools to track and report the changes on their followers. Similar to atheists, they are interested in scientific sources such as n-scientific.org. Tanweeris prefer to access and interact with mix of official and non-official, Arabic and non-Arabic news sources. However, they prefer sources related to traditional newspaper such as alqabas.com, thenewkhalij.news, alghadeer.tv, aljarida.com and alhudood.net. Also, they use Iranian news sources such as mojahedin.org, Iraqi news source such as alsumaria.tv and alghadeer.tv, and one non-arabic source ansa.it. 
This supports our observation that Tanweeri are interested in challenging cultural restrictions in Arab societies and interacting with other cultures as inspiration for reform.

\section{Discussion, Conclusion and Future Work}\label{sec:conclusion}

In this study, we shed some light on a neglected, but important topic, online discussion of Atheism in the Arab world. While our analysis is mostly descriptive and quantitative, we believe that it provides valuable insights about the atheist community in the Arab world and how they interact with other online users, which should provide a solid baseline for future work. %
Our analysis to the most active 434 Arab users on Twitter discussing atheism shows that there is a large discussion of the topic online between mainly three groups: 1) users promoting atheism and argue against religion; 2) users who are refuting atheism and its arguments, and 3) users who does not explicitly deny religions but asking for reform of them.
Our findings shows that a lot of the discussion about atheism in the Arab world includes the situation in the Middle East. Atheists focus more on the rights of some groups in the Arab world, such as violence against women. Theists discuss more the national challenges facing the society. Tanweeris were found to show more solidarity with their governments while criticising Islamic groups and their interpretation of Islam.
We observed that Arab atheists are willing to communicate with foreign cultures such as Western news sources, TV shows, and world wide organisations. 
Tanweeris interact more with traditional news sources such as newspapers, and discuss non-religious content. Theists were found to reference a lot of Islamic content in their tweets.

In future work, we hope to replicate this study with Theist and Tanweeri seed lists, in order to obtain a more rounded picture of Arab religious discourse online. We also plan to investigate the network dynamics and the  directions of interaction and links in more depth. Finally, we may consider building a classifier that determines the current position of a given Twitter account and tracks potential changes over time, but that requires being mindful of potential ethical implications, given that atheism is illegal in several Arab countries.

We hope that this study will motivate applying much deeper analysis to this sensitive topic in the future.

\section*{\ackname}
We would like to thank Bridges Foundation for their support by providing the initial seeding list of accounts that we later reviewed and filtered. In addition, we are highly thankful for Dr. Jacob Copeman, an expert anthropologist in Atheism and non-religion, for his valuable feedback and suggestions for our study.

\bibliographystyle{splncs04}
\bibliography{mybibliography}

\newpage
\centering \appendixname{}
\begin{table}[H]
\centering
\scriptsize
\caption{The highest occurrence of hashtags used by all groups}
\label{tbl:allfreqhastags}
\begin{tabular}{@{}lrrr@{}}
\hline
\centering Hashtag (Translation) & Atheistic & Theistic & Tanweeri \\ \hline
\<عقلانيون> (Rationalists) & 28048 & 16743 & 2377 \\
\<السعوديه> (Saudi Arabia) & 2274 & 1018 & 453 \\
\<داعش> (ISIS) & 972 & 2001 & 86 \\
\<صناعه المحاور> (Creating Almohawer (interlocutor)) & 938 & 1159 & 123 \\
\<الاسلام> (Islam) & 888 & 716 & 75 \\
\<ايران> (Iran) & 514 & 648 & 96 \\
\<سوريا> (Syria) & 543 & 611 & 67 \\
\<مصر> (Egypt) & 663 & 464 & 81 \\
\<اليمن> (Yemen) & 690 & 347 & 71 \\
\<تنوير> (Tanweer) & 836 & 50 & 117 \\
\<قطر> (Qatar) & 351 & 362 & 103 \\
\<عقلانيون جدد> (New rationalists) & 437 & 678 & 19 \\
\<نظريه التطور> (The theory of evolution) & 613 & 454 & 41 \\
\<ستيفن هوكينج> (Stephen Hawking) & 625 & 194 & 94 \\
\<ماجد مانع عمير معنف زوجته> (Majed Mana' Omair oppresse his wife) & 606 & 69 & 123 \\
\<اسرائيل> (Israel) & 552 & 191 & 99 \\
\<العراق> (Iraq) & 460 & 446 & 47 \\
\<حقيقه> (Truth) & 527 & 382 & 52 \\
\<محمد بن سلمان> (MBS) & 264 & 88 & 129 \\
CEDAWSaudi & 251 & 105 & 122 \\
\<الصحوه> (Sahwa) & 293 & 47 & 114 \\
\<النرد الابراهيمي> (Abrahamic dice) & 319 & 3 & 120 \\
\<الكويت> (Kuwait) & 271 & 197 & 80 \\
FreeRaif & 427 & 49 & 84 \\
\<القدس> (Jerusalem) & 279 & 358 & 28 \\
\<نقد الموروث> (Criticise ancestral) & 418 & 58 & 72 \\
\<رمضان> (Ramadan) & 471 & 128 & 47 \\
\<القدس عاصمه فلسطين الابديه> (Jerusalem is the eternal capital of Palestine) & 226 & 352 & 23 \\
Free Sherif Gaber & 529 & 10 & 54 \\
\<سعوديات نطلب اسقاط الولايه> (Saudi women demand dropping of guardianship) & 258 & 83 & 66 \\
\<الرياض> (Riyadh) & 206 & 112 & 64 \\
\<اسقاط الولايه> (Dropping of guardianship) & 235 & 149 & 51 \\ \hline
\end{tabular}
\end{table}

\newpage

\begin{table}[H]
\centering
\scriptsize
\caption{The most frequent hashtags used by each group}
\label{tbl:aafreqhastags}
\begin{tabular}{@{}lrrr@{}}
\hline
\centering Hashtag (Translation) & Atheistic & Theistic & Tanweeri \\ \hline
\<تجاره الوهم> (Trades of illusion) & 859 & 5 & 40 \\
ExMuslim & 719 & 60 & 51 \\
Atheism & 614 & 72 & 27 \\
\<حقيقه التطور> (Evolution is a fact) & 609 & 8 & 29 \\
\<الحريه لشريف جابر> (Free Sharif Jaber) & 533 & 66 & 88 \\
Atheist & 481 & 23 & 29 \\
Tunisie & 474 & 5 & 0 \\
SaveDinaAli & 430 & 5 & 18 \\
\<غرد بصوره> (Tweet a picture) & 369 & 46 & 29 \\
\<نطالب بحريه الاعتقاد> (We demand freedom of belief) & 369 & 0 & 112 \\
\<رائف بدوي> (Raif Badawy) & 361 & 16 & 8 \\
\<علوم> (Science) & 316 & 19 & 30 \\
\<جمعه مباركه> (Blessed Friday) & 313 & 106 & 43 \\
\<القران بالصور> (Quran in pictures) & 293 & 1 & 3 \\
\<ترامب> (Trump) & 281 & 103 & 12 \\
\<ملحد> (Atheist) & 166 & 1214 & 6 \\
\<الالحاد> (atheism) & 129 & 473 & 6 \\
\<مذبحه الاطفال في افغانستان> (Afgan children's massacre) & 16 & 449 & 9 \\
\<دعاء> (Pray) & 2 & 403 & 0 \\
\<انتشارالاسلام> (Spread of Islam) & 0 & 348 & 0 \\
\<حقيقه النصرانيه> (The fact of Christianity) & 7 & 345 & 0 \\
\<يسوع> (Jesus) & 30 & 316 & 1 \\
\<فلسطين> (Palestine) & 165 & 312 & 35 \\
\<الكتاب المقدس> (Bible) & 10 & 297 & 0 \\
\<الازهر قادم> (Al-Azhar is coming) & 11 & 280 & 1 \\
Quran & 42 & 266 & 2 \\
\<الملاحده> (Atheists) & 10 & 262 & 0 \\
\<شرعنه التصهين> (Legalization of zionization) & 0 & 230 & 0 \\
\<حساب الدفاع عن الوحيين بالحجج والبراهين> \\(Defence of Quran and Sunnah by arguments and proofs) & 0 & 201 & 0 \\
\<الاحتلال الهاشمي> (Hashemite occupation -over Yemen-) & 1 & 0 & 236 \\
\<لا تسرق يا عايض> (O Ayed, Do not steal) & 43 & 3 & 140 \\
\<احاديث ضعيفه استغلتها الصحوه> (Weak Hadith employed by sahwa) & 142 & 32 & 85 \\
\<تنظيف المدارس من السروريات> (Cleaning schools from Sururi women) & 64 & 33 & 84 \\
\<بوح ناي> (Flute's revelation) & 13 & 1 & 78 \\
\<ولي العهد> (Crown prince) & 103 & 35 & 75 \\
\<النصر> (Al-Nassr FC) & 18 & 21 & 65 \\
CBS \<ولي العهد علي قناه> (Crown prince on CBS) & 110 & 29 & 64 \\
\<بدل عبوديه للمراه السعوديه> (Slavery allowance for Saudi women) & 26 & 8 & 63 \\
\<كيف نجوت من الصحوه> (How I survive from Sahwa) & 122 & 35 & 59 \\
\<اب يرفض خروج ابنته من السجن> (Refused to release his daughter) & 114 & 66 & 54 \\
\<قررت البس عبايه علي الراس> (I decided to wear it on my head) & 85 & 40 & 54 \\
\<المطلق العبايه غير الزاميه> (Almutlaq Abaya is not obligatory) & 137 & 48 & 52 \\
\<الاخوان> (Brothers) & 57 & 21 & 49 \\
\<اليمن> (Yemen) & 136 & 40 & 47 \\
\hline
\end{tabular}
\end{table}

\begin{figure}[t]
 \centering
 \includegraphics[width=0.9\linewidth]{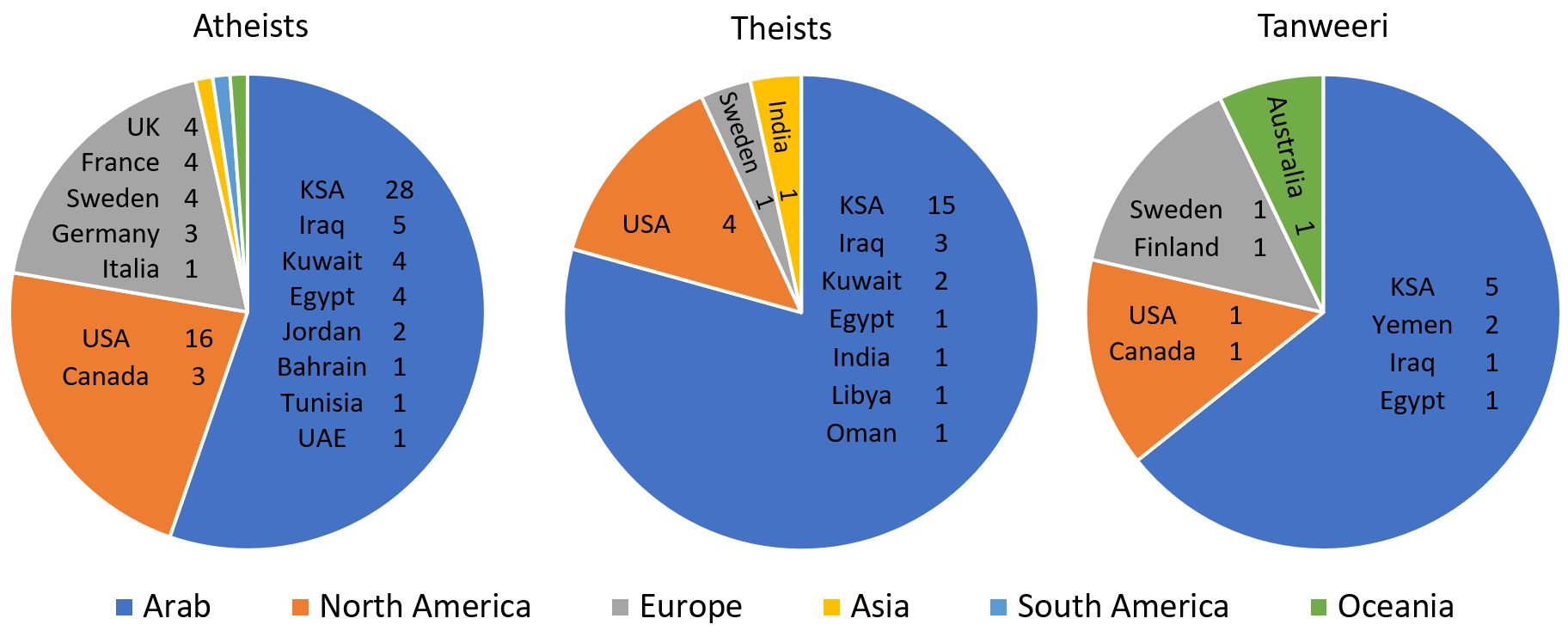}
 \vspace{-0.2cm}
 \caption{Identified Locations for Atheist Group}
 \label{fig:locationsData} 
\end{figure}

\begin{table}
\caption{The 20 most frequent Domains that are used by Each class.}
\vspace{-0.2cm}
\label{tbl:freqDomains}
\begin{minipage}[b]{0.312\textwidth}
\label{tbl:aafreqDomains}
\begin{tabular}{@{}|p{3cm}r|@{}}
\hline
\centering Atheists & \\ \hline
Domain & Freq. \\ \hline
ask.fm & 901 \\
wearesaudis.net & 596 \\
goodreads.com & 439 \\
ibelieveinsci.com & 273 \\
dw.com & 213 \\
al-eman.com & 198 \\
atheistrepublic.com & 170 \\
ahewar.org & 163 \\
atheistdoctor.com & 154 \\
dorar.net & 135 \\
imdb.com & 86 \\
libral.org & 62 \\
linkis.com & 61 \\
arabatheistbroadcasting & 59 \\
dkhlak.com & 58 \\
iqtp.org & 57 \\
syr-res.com & 50 \\
bassam.nu & 45 \\
friendlyatheist.patheos & 33 \\
mustafaris.com & 28 \\
 \hline
\end{tabular}
\end{minipage}
\begin{minipage}[b]{0.321\textwidth}
\label{tbl:pafreqDomains}
\begin{tabular}{@{}|p{3cm}r|@{}}
\hline
\centering Theists &  \\ \hline
Domain & Freq. \\ \hline
du3a.org & 11768 \\
d3waapp.org & 2807 \\
alathkar.org & 1402 \\
kaheel7.com & 375 \\
almohawer.com & 327 \\
sabq.org & 175 \\
unfollowspy.com & 151 \\
bayanelislam.net & 139 \\
antishobhat.blogspot & 129 \\
kutub-pdf.net & 122 \\
spa.gov.sa & 109 \\
i.imgur.com & 108 \\
ncbi.nlm.nih.gov & 97 \\
justpaste.it & 96 \\
7asnat.com & 90 \\
quran.to & 82 \\
survey-smiles.com & 82 \\
cnsnews.com & 75 \\
alkulify.blogspot & 68 \\
estigfar.co & 56 \\
\hline
\end{tabular}
\end{minipage}
\begin{minipage}[b]{0.32\textwidth}
\label{tbl:abfreqDomains}
\begin{tabular}{@{}|p{3cm}r|@{}}
\hline
\centering Tanweer & \\ \hline
Domain & Freq. \\ \hline
fllwrs.com & 323 \\
crowdfireapp.com & 134 \\
eremnews.com & 63 \\
alqabas.com & 59 \\
thenewkhalij.news & 57 \\
8bp.co & 43 \\
n-scientific.org & 35 \\
alghadeer.tv & 21 \\
maktaba-amma.com & 21 \\
alarab.co.uk & 15 \\
telegra.ph & 12 \\
aljarida.com & 12 \\
alhudood.net & 11 \\
dr-alawni.com & 9 \\
alsumaria.tv & 9 \\
ansa.it & 8 \\
arabic.mojahedin.org & 6 \\
arabketab4u.blogspot & 6 \\
marebpress.net & 5 \\
emaratalyoum.com & 5 \\
\hline
\end{tabular}
\end{minipage}
\end{table}

\end{document}